\documentclass[runningheads]{llncs}

\usepackage[year=2026,ID=5]{eccv}
\usepackage{eccvabbrv}

\usepackage{lmodern}
\usepackage[T1]{fontenc}
\usepackage{amsmath,amssymb}
\usepackage{booktabs}
\usepackage{graphicx}
\usepackage{hyperref}
\usepackage{xcolor}
\usepackage{xspace}
\usepackage{cite}
\usepackage{hyperref}

\begin{document}

\title{Soft Redaction of Image Provenance via \\Zero-Knowledge Proofs}

\author{Muhammad Awan\inst{1} \and John Collomosse\inst{1,2}}

\titlerunning{Soft Redaction of Image Provenance via ZKPs}
\authorrunning{M.~Awan and J.~Collomosse}

\institute{University of Surrey, Guildford GU2 7XH, UK \and Adobe Research, San Jose CA 95110, USA}

\maketitle

\begin{abstract}
Content provenance standards, such as C2PA, are increasingly used to attach signed records of origin, editing history, and rights to digital images. However, provenance transparency can conflict with privacy -- assertions that strengthen trust in an image may also reveal sensitive information about the creator or capture context. We propose \emph{soft redaction} for image provenance: a mechanism that replaces sensitive provenance assertions with zero-knowledge proofs (ZKPs) of selected properties over hidden data. Our work focuses on distance proofs. We first show how location assertions can support proofs of proximity to a public reference point, using Chebyshev polynomial approximations within the ZKP proof circuit. We then extend the approach to $\ell_2$ distance proofs over biometric embeddings, enabling privacy-preserving claims related to likeness to help enforce personality rights with images. Finally, we apply the same distance-proof construction to perceptual hashes (visual fingerprints), supporting an anti-spoofing use case in watermark-based recovery of stripped provenance metadata. Our results demonstrate that ZKPs over image provenance can provide practical soft-redaction capabilities, compatible with C2PA, that may be constructed in seconds and verified in milliseconds.
\end{abstract}

\section{Introduction}

The rise of generative AI has led to increased societal expectations around media transparency --  how an image was created, by whom or what process, and how it has been modified. These expectations are driven by concerns around content authenticity, and more recently questions of creator consent in the re-use of content. Such concerns have driven significant interest in {\em media provenance}: the recording and communication of information about the origin, history, and rights associated with digital media, such as images.

The Coalition for Content Provenance and Authenticity (C2PA; ISO/DIS 22144 \cite{c2pa}) is an open cross-industry standard for describing media provenance in asset metadata, and is increasingly adopted in the image modality. Many manufacturers embed it at capture time in high-end cameras (\eg Leica M11-P, Nikon Z9, Canon EOS R5 Mark II) and consumer mobile (\eg Google Pixel 8+) devices, as do generative AI platforms (\eg Gemini, ChatGPT) and creative  tools (\eg Adobe Photoshop).   The purpose of C2PA metadata (referred to as a `C2PA manifest') is to carry cryptographically signed provenance facts (called `assertions') about the asset. These assertions may describe who created the asset, where and how it was created or edited, and which other assets were used in its production. Such referenced assets, known as `ingredients', may themselves carry C2PA manifests, allowing manifests to link to one another, forming a directed provenance graph. For example, as a photograph moves through the content supply chain, it may be transformed by editing tools, with each step adding a further manifest documenting the actions performed. The end consumer of the media can therefore be provided with a history of the asset, placing them in a more informed position to assess its authenticity and origin.  More recently, C2PA manifests have also been used to attach creator preference signals (\eg AI opt-out \cite{decorait}) and licensing contracts \cite{ekila, contentarcs, Collomosse-IEEECGA-2026} to assets.

One concern that has been raised around C2PA is privacy and the potential for provenance metadata to be misused \cite{c2paprivacy}. This is mitigated through user control over which assertions to include and, importantly, through the ability to redact assertions at any point in the content supply chain. For example, a photojournalist operating in a conflict zone may wish to redact their location, or to include it initially but ask a downstream news publisher to perform the redaction before publication. However, this introduces a fundamental tension in that redaction can also weaken the evidentiary value of the provenance record.  It may be sufficient to prove that a photograph was taken in a particular country, without revealing precisely where. Yet, such an ability to perform a partial or `soft redaction' -- replacing an assertion with a verifiable claim about its value rather than the value itself -- is not yet available within provenance standards.

In this paper we explore the potential for performing soft redaction using zero-knowledge proofs (ZKPs) constructed over assertion data, including visual data derived from an image such as biometric identifiers \cite{deng2019arcface,schroff2015facenet,adaface,elasticface}, and perceptual hashes (`fingerprints') \cite{Black-WMF-2021,simprov,sscd}.  ZKPs enable relationships (such as inequalities or distances) to be proven over a secret value (called a `witness') without disclosing the witness to the verifier \cite{thaler2022pazk}.  For example, a ZKP can allow a third party to verify that a person is over 18 years of age without requiring that person to disclose their date of birth. Although ZKPs have been recently explored for proving properties of image pixel transformations~\cite{zkimg,veritas}, to our knowledge no prior work applies them to C2PA provenance assertions.   We make three technical contributions:

\noindent {\bf 1. Location privacy.} We demonstrate an initial proof-of-concept for ZKPs in C2PA using location metadata, enabling proof that an image was captured within a given distance of a public reference point without revealing the precise GPS coordinate. We implement a Chebyshev approximation of the required trigonometric functions, and characterise the accuracy-efficiency trade-off.

\noindent {\bf 2. Likeness provenance.} We extend this approach to high-dimensional assertion data derived from the image. We construct an $\ell_2$ distance proof for biometric embeddings, enabling an image-derived descriptor to be checked against a hidden registered likeness descriptor without disclosing the registered descriptor itself. This supports provenance claims around personality rights; an emerging use case \cite{personalityrights}.

\noindent {\bf 3. Fingerprint-based recovery.} We show that the same $\ell_2$ proof circuit also supports privacy-preserving claims over perceptual image fingerprints. This enables soft-redacted assertions about visual similarity, which we apply as an anti-spoofing measure for watermark-based recovery of provenance -- a common way to mitigate the stripping of C2PA metadata on content platforms \cite{collomosse_authenticity_2024}.

\section{Related Work}
\label{sec:related}

\paragraph{Media provenance} has become increasingly important in a decentralized information sphere where content --- such as news imagery --- originates from a diverse ecosystem of creators rather than a small number of institutional providers. This marks a shift from institutionally mediated trust toward technologically mediated trust~\cite{Bui-CVPRWS-2019}. ARCHANGEL~\cite{Collomosse-DocEng-2018} was among the earliest explorations of this shift, combining visual fingerprinting, cryptographically signed metadata, and blockchain records to protect the integrity of media in sovereign public archives~\cite{Bui-CVPRWS-2019,Bui-TMM-2020}. ARCHANGEL was subsequently extended to the news domain through the {\em Angel's Wing} browser extension~\cite{AngelsWings}, which searched a blockchain for provenance records associated with photographs. Blockchain-based approaches have also been proposed for storing media provenance metadata~\cite{AMP-2021}, and  establishing the provenance of photographs in insurance evidentiary contexts~\cite{Gipp-2016}.

Cryptographic protection of provenance metadata is now central to the Coalition for Content Provenance and Authenticity (C2PA) open standard~\cite{c2pa}. C2PA aims to persist provenance data from ``glass to glass'': from a C2PA-enabled capture device, through the content supply chain, to the final consumer. In practice, however, many content platforms (\eg social media) strip C2PA metadata, meaning that signed metadata alone is often insufficient to communicate provenance reliably~\cite{collomosse_authenticity_2024}. This motivates complementary content re-identification technologies, such as invisible watermarking~\cite{trustmark}, which inject signals into the asset that can be used to recover lost provenance information from a database.  Security vulnerabilities inherent to watermarks (such as the ability to strip and spoof them) further motivate the use of visual fingerprinting to reinforce them.  There are established design patterns for combining these `three pillars of provenance'  (signed metadata, watermarking and fingerprinting) to help provenance survive practical content supply chains~\cite{collomosse_authenticity_2024}.  We use ZKPs not only to soft-redact assertions within C2PA, but also to improve security properties of fingerprinting when used for watermark-based recovery.

\paragraph{Zero-knowledge proof (ZKP) systems} initially relied on encrypting the secret (witness) and proving
properties over the ciphertext~\cite{paillier1999}. In such systems proof and ciphertext sizes both scale with the secret
dimension, making them infeasible for high-dimensional visual embeddings.  Modern polynomial proof systems (SNARKs~\cite{groth16,plonk}) arithmetise
a predicate as polynomial constraints over a finite field; the prover commits
to a witness polynomial and responds to a random evaluation challenge.
This yields constant-size proofs regardless of witness
dimension and sub-second verification, aligned to our goals of computing proofs over C2PA assertion data.
Groth16~\cite{groth16} achieves the smallest proofs at the cost of a
per-circuit trusted setup; PLONK~\cite{plonk} uses a universal setup that
serves all circuits up to a size bound, avoiding per-predicate ceremonies.
Bulletproofs~\cite{bulletproofs} require no setup but have $O(n)$
verification, which is prohibitive for publishers processing large media
catalogues.
We adopt PLONK throughout having empirically compared all
three paradigms on the same predicate (see Section~\ref{sec:zkpchoice}).

\paragraph{ZKP over image transformations} were first explored in PhotoProof~\cite{photoproof}, which established the concept of proof-carrying data for visual content but was limited to small images.  ZK-IMG~\cite{zkimg}, and later VerITAS~\cite{veritas,hyperveritas}, used ZKP to counter visual disinformation using zk-SNARKs.  These works define ZKP over a small set of edit classes (crop, rotate, brightness adjust) and can prove that an image resulted from applying them to a camera-signed original.  By contrast, we explore the complementary use of ZKP to perform soft-redaction of C2PA assertion data itself --- including for biometric and content verification use cases on multi-dimensional visual feature embeddings --- without any reference to the image pixels themselves.  Other visual uses of zk-SNARKs include enabling a prover to show that a visual embedding was produced by a specific model~\cite{zkml}.  ZKP for GPS location was explored using floating-point SNARK circuits in~\cite{zklocation}.  We instead use integer fixed-point arithmetic with a degree-5 Chebyshev approximation of the Haversine (great-circle) formula.

\paragraph{Facial biometrics} are a well-studied field that have converged on a common paradigm: a deep residual network
(IR50 or IR100~\cite{deng2019arcface}) trained with an angular margin loss produces a compact $\ell_2$-normalised embedding.
ArcFace~\cite{deng2019arcface} introduced additive angular margin as the standard training objective; subsequent works refined the margin schedule: AdaFace~\cite{adaface} adapts the margin to image quality, while ElasticFace~\cite{elasticface} relaxes the fixed-margin constraint with an elastic penalty, training separate Arc- and Cosine-margin variants. FaceNet~\cite{schroff2015facenet} uses an InceptionResnet-V1 backbone with
triplet loss, producing 512D embeddings pre-trained on VGGFace2.
All models produce unit-normalised embeddings for inference, matching via
$\ell_2$ distance -- the predicate our ZKP computes directly.

\section{Soft Redaction of C2PA Manifests}
\label{sec:foundationmath}

A C2PA manifest is a cryptographically signed (X.509) metadata structure containing a collection of provenance facts (assertions) which may include location or other descriptors and facts about the provenance of an image.  Assertions are stored within an internal assertion store and referenced from the signed claim via hashed URIs. C2PA already supports \emph{hard redaction}, whereby an assertion is removed from the assertion store while its hash $H_a$ remains in
the signed claim, allowing the redaction itself to be audited without revealing the underlying value.

Our approach extends this mechanism to support \emph{soft redaction}. Rather than simply deleting sensitive metadata, the assertion is replaced by a ZKP demonstrating that the hidden value satisfies a chosen predicate. The existing assertion hash provides a cryptographic binding between the original camera-signed assertion and the subsequently attached proof.

When a party in the content supply chain performs soft redaction they

\begin{enumerate}
\item read the assertion value $V$ from the camera-signed manifest;
\item generate a ZKP that $V$ satisfies predicate $P$;
\item hard-redact the original assertion while retaining its hash $H_a$ in the
signed claim; and
\item attach the ZKP as a new assertion within a subsequent C2PA update
manifest.
\end{enumerate}

\begin{definition}[Soft Redaction]
Let $\mathcal{M}$ be a C2PA manifest containing assertion
$a=(V,H_a)$. A \emph{soft redaction} of $a$ with predicate $P$ is the tuple
$(C,\theta,\pi)$ where
$C=\mathrm{commit}(V,r)$,
$\theta$ denotes the public parameters of $P$, and $\pi$ is a
zero-knowledge proof of

\[
\exists\,V,r:
C=\mathrm{commit}(V,r)
\wedge
P(V;\theta)=\top.
\]

The original assertion is hard-redacted and the tuple
$(C,\theta,\pi)$ is stored as a new assertion within an update manifest.
\end{definition}

This paper focuses on distance predicates, which prove that a hidden witness value
$\mathbf{v} \in \mathbb{R}^D$ lies within radius $R$ of a public reference:
$\|\mathbf{v} - \mathbf{v}_{\mathrm{ref}}\|_2 \leq R$.  The remainder of this Section briefly explores a proof of concept for soft redaction of GPS location, before proceeding to reformulate to  $\ell_2$ distance for visual biometric and fingerprinting
use cases in Sections~\ref{sec:biometric}--\ref{sec:fingerprint}. GPS location provides a convenient proof-of-concept because it involves low-dimensional non-visual metadata while still requiring a non-trivial arithmetic approximation. The same predicate formulation subsequently generalises to high-dimensional visual embeddings.


\subsection{Approximating Location Distance}
\label{sec:location}

We first consider soft redaction of GPS coordinates, one of the most common privacy-sensitive assertions contained within C2PA manifests. Rather than revealing the precise capture location, the objective is to prove that the hidden coordinate lies within a specified distance of a public reference point.

The geodesic (great-circle) distance between two latitude--longitude pairs is given by the Haversine formula

\[
d
=
2R_\oplus
\arcsin
\left(
\sqrt{
\sin^2\!\left(\frac{\Delta\phi}{2}\right)
+
\cos\phi_1
\cos\phi_2
\sin^2\!\left(\frac{\Delta\lambda}{2}\right)
}
\right),
\]

where $R_\oplus$ denotes the Earth's radius, $\phi$ latitude and $\lambda$
longitude.

Direct evaluation is expensive within arithmetic ZKP systems because transcendental functions expand into large constraint systems. Rather than
evaluating the distance explicitly, we reformulate the predicate

\[
d \le R
\]

as an equivalent comparison over the intermediate Haversine accumulator

\[
a
=
\sin^2\!\left(\frac{\Delta\phi}{2}\right)
+
\cos\phi_1
\cos\phi_2
\sin^2\!\left(\frac{\Delta\lambda}{2}\right),
\]

giving

\[
a
\le
a^\star
=
\sin^2\!\left(\frac{R}{2R_\oplus}\right).
\]

The remaining
$\sin$ and $\cos$ terms are approximated using degree-$d$ Chebyshev
polynomials fitted over $[-\pi/2,\pi/2]$, with approximation error
$\epsilon_d$ absorbed into the acceptance threshold. All computations are performed in fixed-point arithmetic over the circuit field.

Table~\ref{tab:cheby_empirical} reports the empirical $p_{99}$ absolute error of the
approximation evaluated over 20,000 randomly sampled coordinate pairs spanning
city, regional, country and continental scales. Degree-5 Chebyshev
polynomials provide the best practical compromise between circuit complexity
and approximation accuracy, with $p_{99}$  errors of only 36\,m, 214\,m, 1.1\,km and
7.6\,km respectively.

\begin{table}[h]
\centering
\caption{Comparing distance error (metres) by distance approximation
method for various geographic scales.}
\label{tab:cheby_empirical}
\resizebox{\columnwidth}{!}{%
\begin{tabular}{@{}lrrrr@{}}
\toprule
Method & City (50\,km) & Region (300\,km) & Country (1500\,km) & Continent (8000\,km) \\
\midrule
Equirectangular  &        0.2 &         51 &      7,766 & 3,173,252 \\
Chebyshev-3          &    1,353   &      8,104 &     43,542 &   405,267 \\
Chebyshev-5          &       36   &        214 &      1,106 &     7,604 \\
Chebyshev-7          &        0.1 &          0.8 &        6.8 &        62 \\
Chebyshev-9          &     0.004  &        0.02 &        0.1 &       0.6 \\
\bottomrule
\end{tabular}%
}
\end{table}

Implementing the Chebyshev-5 approximation as a PLONK circuit over BN128 requires
three in-circuit trigonometric approximations and compiles to
334 constraints. On a commodity CPU (MacBook M3 Max, 48\,GB), proof
generation required 0.64\,s on average over 100 randomly generated C2PA
manifests, while verification required only 222\,ms. The resulting proof
occupies 768\,B, comfortably within typical C2PA manifest sizes. Since the
current C2PA specification defines no standard assertion type for
zero-knowledge proofs, we store the proof as a custom assertion in our prototype.

\subsection{Choice of ZKP Algorithm}
\label{sec:zkpchoice}

To select a proving system suitable for C2PA soft redaction we compare
Bulletproofs~\cite{bulletproofs}, Groth16~\cite{groth16} and
PLONK~\cite{plonk} using the same 73-constraint equirectangular predicate
($R=150$\,km, eight sample points) on the same hardware, using \texttt{snarkjs}~0.7.

Bulletproofs require no trusted setup and produce an
$O(\log n)$ proof (1,056\,B for $n=36$ bits), but verification scales
linearly with circuit size, requiring approximately 1.5\,s for this simple
predicate. Groth16 produces the smallest proofs (256\,B) and fastest proving
time (0.29\,s), but requires a new trusted setup ceremony for every predicate.
PLONK provides a universal structured reference string (SRS), allowing a single
setup to support all circuits within a size bound while maintaining constant
verification time (220\,ms) and a fixed proof size of 768\,B. Although larger
than Groth16 proofs, this overhead is negligible relative to typical C2PA
manifest sizes. We therefore adopt PLONK throughout the remainder of the paper.

\section{Soft Redaction of Facial Biometrics}
\label{sec:biometric}

Provenance standards such as C2PA focus primarily on communicating the creation
history of an asset. Recent works such as ContentARCS \cite{contentarcs} and ZOETROPE~\cite{Collomosse-IEEECGA-2026}
has extended this framing to machine-readable rights expressions, enabling
creators to specify and monetise downstream reuse of visual assets. However, an
equally important class of rights attaches not to the \emph{asset} but to the
\emph{individuals depicted within it}: personality rights, including consent and
compensation for the use of one's likeness. This issue is particularly acute in
AI-generated imagery, where celebrity likeness reuse has already led to
high-profile legal disputes.

Enforcing such rights at scale requires two capabilities. First, a depicted person
must be able to assert rights over their likeness, for example through a
decentralised personality-rights registry containing usage permissions and
licensing terms. Second, third parties must be able to determine whether an image
contains that person's likeness, without requiring disclosure of a reusable
biometric template. The natural primitive for this is a face recognition query.
A rights-holder registers a protected biometric descriptor of their likeness
$\mathbf{q} \in \mathbb{R}^{D}$, computed from enrolled images, together with
the terms under which their likeness may be reused. A third party wishing to use
an image runs the same biometric model over that image to obtain an image-derived
descriptor $\mathbf{v} \in \mathbb{R}^{D}$, and must determine whether
$\mathbf{v}$ matches the registered descriptor $\mathbf{q}$. Figure~\ref{fig:exp6_det} (right) shows our implementation of this system.

State-of-the-art face recognition networks, including
ArcFace~\cite{deng2019arcface}, FaceNet~\cite{schroff2015facenet},
AdaFace~\cite{adaface}, and ElasticFace~\cite{elasticface}, produce
unit-normalised $D$-dimensional embeddings ($D=128$--$512$), for which the
 identity-verification predicate is
\[
  \|\mathbf{q} - \mathbf{v}\|_2 \leq R .
\]
Public disclosure of $\mathbf{q}$ is unacceptable: a facial biometric template is
a sensitive and effectively irrevocable credential, its disclosure enables
cross-platform tracking, and it may allow adversaries to infer or spoof the
rights-holder's likeness. The registry thus should support verification
against $\mathbf{q}$ without revealing $\mathbf{q}$ itself.

A soft-redaction proof resolves this tension by replacing the plaintext biometric
descriptor with a zero-knowledge proof of proximity. The registry stores a
soft-redacted biometric assertion for the rights-holder, and a verifier supplies
the image-derived descriptor $\mathbf{v}$ as a public input. The proof establishes
\[
  \pi \;:\; \exists\, \mathbf{q} \;:\;
  \|\mathbf{q} - \mathbf{v}\|_2^2 \;\leq\; R^2
\]
where $\mathbf{q}$ is the hidden registered descriptor and $(\mathbf{v}, R^2)$
are public inputs. The verifier learns only whether the image descriptor lies
within distance $R$ of the registered likeness descriptor; it does not learn the
descriptor itself. If the proof verifies, the user takes relevant steps to honour any relevant terms of use.

\subsection{Biometric comparison ZKP circuit design}

The $\ell_2$ ZKP circuit is implemented using PLONK. Since the predicate is a
Euclidean distance comparison, it reduces to a sum of squared differences in
fixed-point arithmetic and requires no transcendental functions or polynomial
approximations. Embedding components $v_i, q_i \in [-1,1]$ are quantised at scale
$S = 1{,}000$:
\[
  x_i^{(\mathrm{int})} = \lfloor x_i \cdot S \rfloor .
\]
The resulting integer squared distance approximates the real-valued distance as
\[
  \|\mathbf{q}^{(\mathrm{int})} - \mathbf{v}^{(\mathrm{int})}\|_2^2
  \approx
  S^2 \|\mathbf{q} - \mathbf{v}\|_2^2 .
\]
The decision predicate therefore becomes
\[
  \sum_{i=1}^{D}
  \left(q_i^{(\mathrm{int})} - v_i^{(\mathrm{int})}\right)^2
  \leq R^2 S^2 .
\]

The Circom circuit, parameterised by embedding dimension $D$, computes:
\begin{enumerate}
  \item $\delta_i \leftarrow q_i^{(\mathrm{int})} - v_i^{(\mathrm{int})}$
    for $i = 1,\ldots,D$ \quad (linear, zero non-linear constraints);
  \item $s_i \leftarrow \delta_i^2$ for each $i$
    \quad ($D$ non-linear constraints);
  \item $\Sigma \leftarrow \sum_i s_i$
    \quad (linear accumulation);
  \item $\mathrm{slack} \leftarrow R^2 S^2 - \Sigma \geq 0$
    \quad ($\lceil \log_2(R^2 S^2 + 1) \rceil$-bit range check).
\end{enumerate}

For $D = 128$, $S = 1{,}000$, and $R = 0.9$, the threshold is
$R^2 S^2 = 810{,}000 < 2^{21}$, so a 21-bit range check suffices. The compiler
(circom~2.2.3) reports \textbf{279 constraints}: 128 squarings, 21 range-check
bits, and 130 identity constraints wiring the 128-element public image embedding
and threshold into the R1CS.

The public inputs are $\mathbf{v}$, represented as 128 integers, and
$l2sq\_thresh$, giving 129 public signals in total. The private input is
$\mathbf{q}$, represented as 128 integers. All intermediate values fit within
32-bit integer ranges for the thresholds considered, and the circuit requires no
integer division or remainder witnesses. The $\ell_2$ circuit requires fewer constraints than the country-scale
Haversine circuit of Section~\ref{sec:location} (279 vs.\ 334) despite operating
in 128 dimensions. This is because L2 distance is a degree-2 polynomial,
whereas geodesic distance required trigonometric approximation.

\subsubsection{Embedding dimension.}
\label{sec:tradeoffs}

\begin{figure}[h]
\centering
\includegraphics[width=0.9\columnwidth]{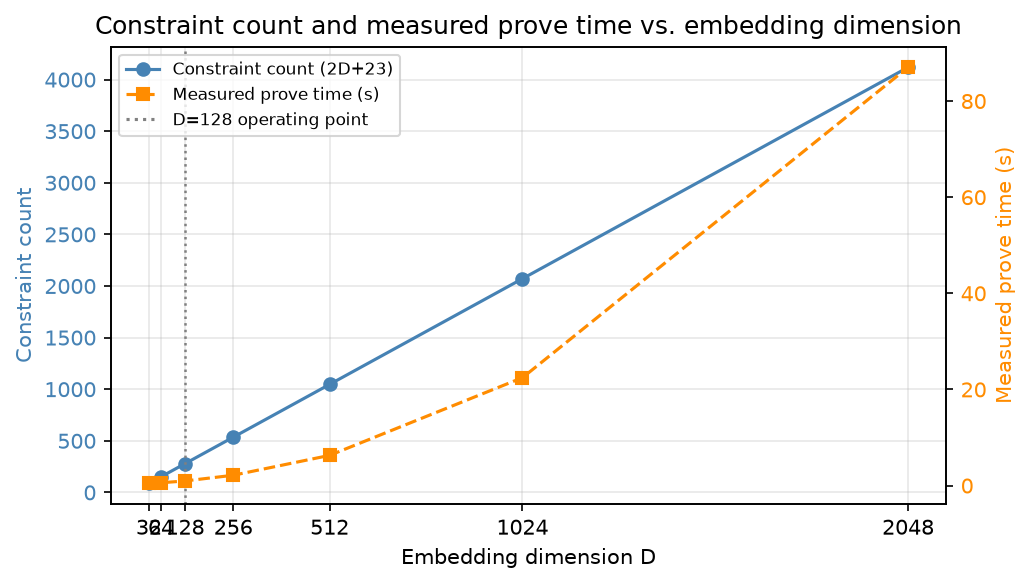}
\caption{Exploring the practiality of ZKP using PLONK.  Proof constraint count (exact formula $2D+23$, left axis) and
measured prove time (right axis; $N=5$ proofs per point, PLONK, M3~Max CPU)
as a function of embedding dimension $D$.
$D\!\leq\!512$: pot12; $D=1024$: pot13; $D=2048$: pot14.
Prove time grows super-linearly as each SRS tier doubling multiplies MSM cost
(${\approx}3\times$ per tier on M3~Max CPU).
The vertical dashed line marks the $D=512$ operating point for biometric descriptors in our work.}
\label{fig:tradeoff_dim}
\end{figure}

Figure~\ref{fig:tradeoff_dim} characterises the effect of embedding
dimension $D$ on circuit cost. Constraint count grows exactly as $2D + 23$
($D$ squarings, 21 range-check bits, and $D+2$ wiring constraints), confirmed
by the Circom compiler at $D=128$. Prove time grows super-linearly with
dimension: measured values are 0.44\,s ($D=32$), 0.58\,s ($D=64$),
0.96\,s ($D=128$), 2.15\,s ($D=256$), 6.28\,s ($D=512$),
22.4\,s ($D=1024$), and 87\,s ($D=2048$). Each SRS tier doubling roughly
triples prove time, as PLONK's MSM operations scale as $O(n \log n)$ in SRS
size. Verify time remains approximately constant at ${\approx}250$\,ms across
all $D$, since it is determined by the PLONK verifier equation count rather
than circuit size. Thus, $D=128$ offers sub-second proving for low-dimensional
descriptors, while $D=512$ remains practical for interactive or offline rights
checks on commodity hardware.

\subsubsection{Recognition threshold.}

The threshold $R$ determines the biometric operating point: lowering $R$
reduces false accepts but increases false rejects, while raising $R$ has the
opposite effect. Importantly, this application-level tradeoff has negligible
effect on circuit complexity. The range-check bit-width is
$b=\lceil\log_2(R^2S^2+1)\rceil$; with $S=1{,}000$, moving across the practical
range of face-recognition thresholds changes $b$ by only a small number of bits.
The verifier can therefore adjust $R$ to match the desired  operating
threshold without recompiling the circuit.

\subsection{Comparative Evaluation of Models}
\label{sec:exp6}

We evaluate the biometric likeness-query circuit on the Labeled Faces in the
Wild (LFW) benchmark~\cite{lfw} using five state-of-the-art face recognition
models. All models produce $D=512$ unit-normalised embeddings; we use the
1{,}047-constraint PLONK circuit with a pot12 SRS
(see Section~\ref{sec:tradeoffs}).

\subsubsection{Dataset.}

LFW provides 6{,}000 pre-formed pairs (3{,}000 same-person, 3{,}000
different-person), allowing EER to be computed per model.
Face images are $125\times 94\times 3$ crops (deep-funnelled alignment).
We apply model-specific preprocessing: ArcFace, AdaFace, and both
ElasticFace variants receive $112\times112$ BGR images normalised to $[-1,1]$;
FaceNet receives $160\times160$ RGB images normalised to $[-1,1]$.
No additional face detection or 5-landmark alignment is applied.

The $D=512$ $\ell_2$ circuit has 1{,}047 constraints ($2D+23$).
Embeddings are quantised at $S=1{,}000$; the threshold
$l2sq\_thresh = \lfloor R^2 S^2 \rfloor$ is set to each model's
empirical EER threshold on LFW.
We prove 5 same-person pairs per model and verify each proof.

\subsubsection{Results and Discussion}

Table~\ref{tab:exp6_models} lists the five models evaluated.
ArcFace~\cite{deng2019arcface}, AdaFace~\cite{adaface}, and the two ElasticFace
variants~\cite{elasticface} use IR50 or IR100 backbones trained on
MS1MV2; all produce $D=512$ embeddings.
FaceNet~\cite{schroff2015facenet} uses InceptionResnet-V1 trained on
VGGFace2.
ElasticFace-Cos optimises a cosine-margin objective and outputs un-normalised
features; we $\ell_2$-normalise before computing distances and before the ZKP
circuit.

\begin{table}[h]
\centering
\caption{Face recognition models evaluated.}
\label{tab:exp6_models}
\begin{tabular}{@{}lllrl@{}}
\toprule
Model & Backbone & Training set & $D$ & Reference \\
\midrule
ArcFace          & IR50        & MS1MV2   & 512 & \cite{deng2019arcface} \\
FaceNet          & InceptionV1 & VGGFace2 & 512 & \cite{schroff2015facenet} \\
AdaFace          & IR50        & MS1MV2   & 512 & \cite{adaface} \\
ElasticFace-Arc  & IR100       & MS1MV2   & 512 & \cite{elasticface} \\
ElasticFace-Cos  & IR100       & MS1MV2   & 512 & \cite{elasticface} \\
\bottomrule
\end{tabular}
\end{table}

\begin{figure}[t!]
\centering
\includegraphics[width=0.52\columnwidth]{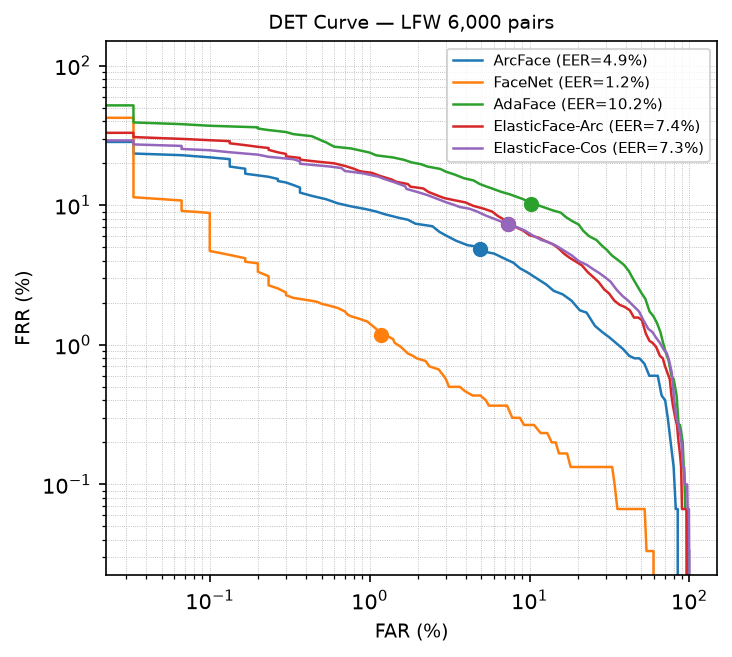}
\includegraphics[width=0.45\columnwidth]{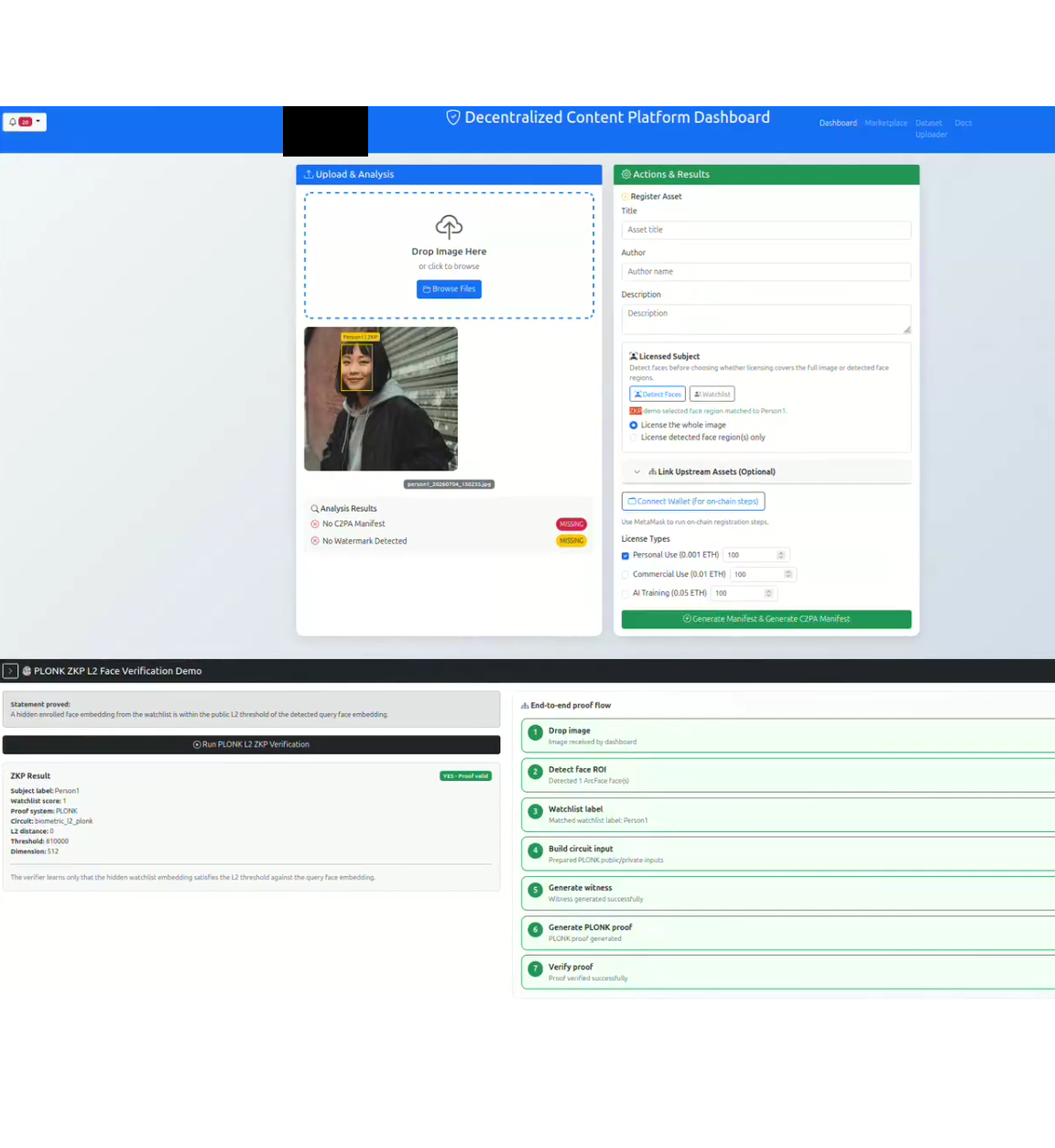}
\caption{Left: Detection Error Trade-off (DET) curves for the five face recognition
models evaluated on LFW. The false accept rate (FAR) is the
percentage of different-person pairs incorrectly accepted as a match. The false reject rate (FRR) is the percentage of same-person pairs
incorrectly rejected. The equal-error rate (EER) is where FAR and FRR become equal. Right: Proof of concept personality rights implementation showing C2PA based licensing, using ZKPs of FaceNet descriptors.}
\label{fig:exp6_det}
\end{figure}

Table~\ref{tab:exp6_results} reports recognition metrics and
Figure~\ref{fig:exp6_det} shows the DET (Detection Error Trade-off) curves and a screenshot of prototype that uses our proposed  approach to lookup licensing metadata for matched likenesses.

\begin{table}[t!]
\centering
\caption{Recognition and ZKP performance on LFW ($D=512$, 6{,}000 pairs).
EER: equal-error rate. $R^*$: L2 threshold at EER.
Acc: verification accuracy at $R^*$.
Prove: mean PLONK prove time (5 same-person pairs, $D=512$, pot12).
Verify: mean PLONK verify time.}
\label{tab:exp6_results}
\begin{tabular}{@{}lrrrrrr@{}}
\toprule
Model & EER (\%) & $R^*$ & Acc (\%) & Prove (s) & Verify (ms) & Proofs OK \\
\midrule
ArcFace (IR50)          & 4.88  & 1.306 & 95.12 & 6.83 & 325 & 5/5 \\
FaceNet (InceptionV1)   & 1.18  & 1.115 & 98.82 & 6.77 & 324 & 5/5 \\
AdaFace (IR50)          & 10.22 & 1.282 & 89.78 & 6.67 & 324 & 5/5 \\
ElasticFace-Arc (IR100) & 7.38  & 1.290 & 92.62 & 6.67 & 309 & 5/5 \\
ElasticFace-Cos (IR100) & 7.28  & 1.298 & 92.72 & 6.62 & 306 & 5/5 \\
\bottomrule
\end{tabular}
\end{table}

Prove time for $D=512$ is approximately 6.7\,s per proof across all models;
verify time is approximately 320\,ms, consistent with the dimension-scaling
measurements in Section~\ref{sec:tradeoffs}. The threshold $R^*$ determines
both the biometric false-accept rate and the size of the set of descriptors that
can satisfy the distance predicate. Lower thresholds make the predicate more
selective, while higher thresholds improve recall at the cost of accepting a
larger neighbourhood around the registered descriptor. In this sense, the
biometric operating point also determines the selectivity of the corresponding
ZKP claim. On a commodity CPU (MacBook M3 Max, 48\,GB), $D=512$ proving takes
6.7\,s, while verification (\ie for a client to check for rights) requires sub-second processing.
\section{Soft Redaction of Visual Fingerprints}
\label{sec:fingerprint}

A practical challenge for C2PA is that content platforms (\eg social media sites)
routinely strip manifest metadata from images. Collomosse and
Parsons~\cite{collomosse_authenticity_2024} describe a three-pillar recovery
pipeline that addresses this problem: an invisible watermark embedded in the
image carries a short identifier payload; a distributed ledger maps this
identifier to the URI of the signed C2PA manifest; and a visual fingerprint
stored inside that manifest performs an anti-spoofing check to confirm that the
retrieved manifest genuinely belongs to the query image, rather than to a
different image onto which the watermark identifier has been maliciously
transferred, or ``spoofed''.

The fingerprint in this pipeline is a compact neural embedding
$\boldsymbol{\phi} \in \mathbb{R}^D$, trained contrastively to be invariant to
non-editorial renditions such as JPEG recompression, minor crops, and brightness
change, while remaining sensitive to semantic content change~\cite{collomosse_authenticity_2024}.
At ingest time, a reference fingerprint $\boldsymbol{\phi}_{\mathrm{ref}}$ is
computed from the original image and stored as a C2PA assertion in the signed
manifest. At query time, a fingerprint $\boldsymbol{\phi}_{q}$ is extracted from
the image under inspection. If
\[
  \|\boldsymbol{\phi}_{q} - \boldsymbol{\phi}_{\mathrm{ref}}\|_2 \leq R ,
\]
then the anti-spoofing check passes and the recovered manifest is accepted as
corresponding to the query image.

Storing $\boldsymbol{\phi}_{\mathrm{ref}}$ in plaintext inside a publicly
retrievable manifest is undesirable on two grounds. First, the fingerprint
descriptor may be a proprietary signal whose exact values are commercially
sensitive, and could be used to reverse-engineer the fingerprinting model or
craft adversarial images that pass the anti-spoofing check. Second, direct
plaintext comparison can leak whether two manifests share a provenance record,
enabling unintended cross-referencing of images. Soft redaction addresses these
concerns by replacing the plaintext reference fingerprint with a zero-knowledge
proof of proximity to a query fingerprint. The verifier learns whether
$\boldsymbol{\phi}_{q}$ lies within distance $R$ of the hidden reference
fingerprint, but does not learn $\boldsymbol{\phi}_{\mathrm{ref}}$ itself. The
same $\ell_2$ circuit used for biometric comparison in
Section~\ref{sec:biometric} applies here without modification.

\subsection{Circuit and Protocol}

In the proposed system, the C2PA manifest stores a soft-redacted reference
fingerprint assertion rather than the plaintext descriptor
$\boldsymbol{\phi}_{\mathrm{ref}}$. The manifest, or a manifest-linked service,
retains a cryptographic binding to $\boldsymbol{\phi}_{\mathrm{ref}}$ through the
C2PA assertion-hash mechanism, while withholding the descriptor value itself.

At query time, a verifier such as a browser extension or content platform
extracts a query fingerprint $\boldsymbol{\phi}_q$ from the image under
inspection. The prover, typically the entity that originally signed the manifest
or a delegated provenance-recovery service, supplies the hidden reference
fingerprint $\boldsymbol{\phi}_{\mathrm{ref}}$ as the private witness and
generates a PLONK proof
\[
  \pi \;:\; \exists\, \boldsymbol{\phi}_{\mathrm{ref}} \;:\;
  \|\boldsymbol{\phi}_q - \boldsymbol{\phi}_{\mathrm{ref}}\|_2^2 \leq R^2 .
\]
Here, $\boldsymbol{\phi}_q$ and $R^2$ are public inputs, while
$\boldsymbol{\phi}_{\mathrm{ref}}$ remains private. The verifier learns only
whether the anti-spoofing predicate holds, not the value of the reference
fingerprint. The descriptor is therefore protected both from casual inspection
of the manifest and from adversaries who might use knowledge of
$\boldsymbol{\phi}_{\mathrm{ref}}$ to craft images that pass the spoofing check.

Operationally, this converts watermark-based recovery into a privacy-preserving
challenge-response protocol. The watermark and ledger recover the candidate
manifest, while the ZKP establishes that the recovered manifest is visually bound
to the query image. If the proof verifies, the recovered C2PA manifest is accepted
as corresponding to the image under inspection; otherwise, the recovered manifest
is rejected as a potential spoof -- as the watermarked identifier upon which the lookup was performed was likely transferred onto a non-matching image.

\subsection{Evaluation on MIRFLICKR-25k}

\subsubsection{Dataset}
We evaluate across $N{=}1{,}000$ images sampled randomly from
MIRFLICKR-25k~\cite{mirflickr}, a standard Creative Commons benchmark for
copy detection and image retrieval.  The images are ingested into the open-source implementation of ZOETROPE~\cite{Collomosse-IEEECGA-2026}, which adds a C2PA manifest and embeds a unique identifier into each image using the TrustMark invisible watermark~\cite{trustmark}.
For each watermarked image we generate five benign transforms: JPEG recompression
at $q{=}75$ and $q{=}50$, 5\% centre crop, $+10\%$ brightness, and $+10\%$
contrast adjustment, yielding 5{,}000 benign query pairs in total. Each transformed image was verified to remain decodable by the ZOETROPE watermark pipeline, confirming that the C2PA manifest remains recoverable after transformation. Watermark spoofing attacks are simulated by re-watermarking images using identifiers from
two randomly selected distinct images from the set, yielding 2{,}000
attack pairs.

\subsubsection{Fingerprinting algorithms.}

The decision threshold $R$ is set per descriptor to $1.2\times$ the maximum
observed benign $\ell_2$ distance across all reference images.

We compare four descriptors spanning a range of training specificity.
\emph{ResNet18-ImageNet} (RN18; 512D)~\cite{he2016deep} uses a standard
ImageNet-supervised ResNet18 with global-average pooling and $\ell_2$
normalisation, with no copy-detection or fingerprinting training.
\emph{DINO ViT-S/8} (384D)~\cite{dino} is a self-supervised vision transformer
trained with the DINO objective on ImageNet, providing stronger semantic features
than a supervised classifier but still without explicit copy-detection training.
\emph{SimProv} (256D) is the contrastively trained component of the image comparator network (ICN) model trained
 for photographic identity verification~\cite{Black-WMF-2021},
extended to use global-average pooling over local spatial features\cite{simprov}.
\emph{SSCD}~\cite{sscd} (512D) is a ResNet-50 trained with the DISC self-supervised
copy-detection objective, producing unit-normalised embeddings with strong
invariance to photographic rendition while remaining discriminative for semantic content.

\subsubsection{Results and Discussion.} 
\begin{figure}[t!]
\centering
\includegraphics[width=\columnwidth]{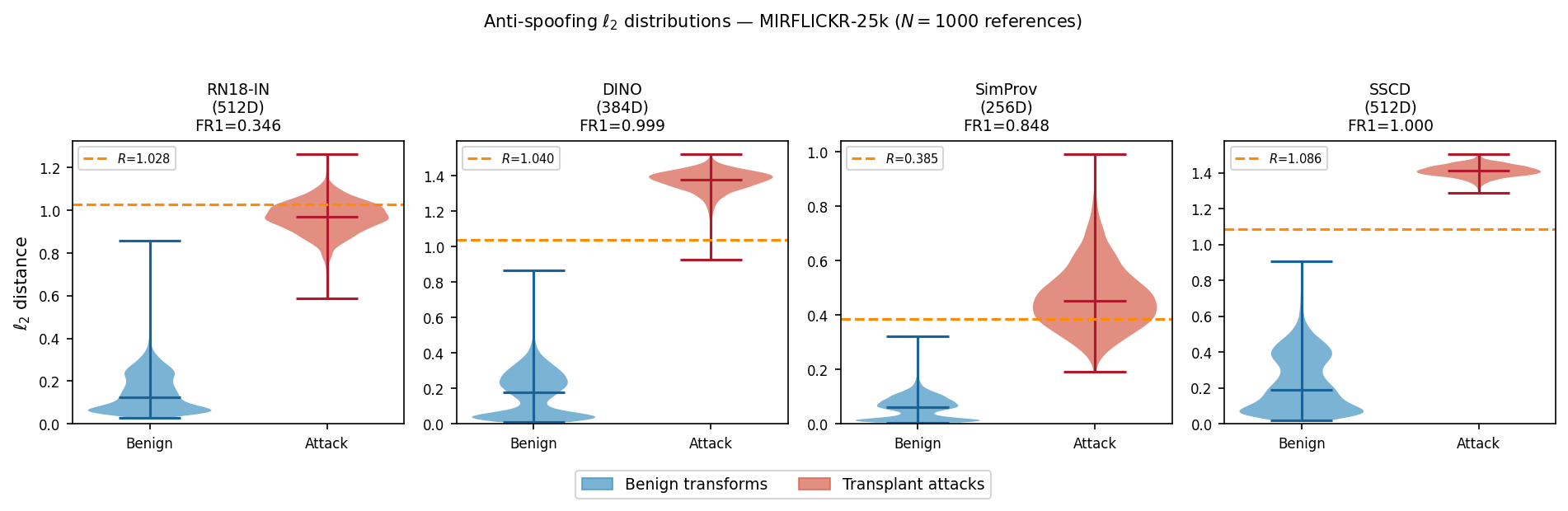}
\caption{$\ell_2$ distance distributions for four descriptors (RN18-IN,
DINO, SimProv, SSCD) across 5{,}000 benign transforms (blue) and 2{,}000
transplant attacks (red) on 1{,}000 MIRFLICKR-25k reference images.
Dashed orange line: decision threshold $R$.
Descriptors ordered left-to-right by training specificity.}
\label{fig:fingerprint_mirflickr}
\end{figure}

Table~\ref{tab:fingerprint_mirflickr} reports aggregate results.
Attack rejection increases sharply with training specificity.
RN18-IN rejects only 418/2{,}000 attacks (20.9\%, FR1\,=\,0.35): generic
ImageNet features are inadequate because semantically similar images share classifier activations.
SimProv rejects 73.6\% of attacks (FR1\,=\,0.85), achieving partial separation but
with a 26\% miss rate due confusable MIRFLICKR pairs.
DINO ViT-S/8, despite no copy-detection training, rejects 1{,}994/2{,}000
attacks (99.7\%, FR1\,=\,1.00): the patch-level self-attention mechanism of ViT
creates instance discriminative features that generalise well to this task.
SSCD achieves perfect rejection (FR1\,=\,1.00). Figure~\ref{fig:fingerprint_mirflickr} shows the full $\ell_2$ distributions;
the benign--attack separation narrows as training specificity decreases.

\begin{table}[h]
\centering
\caption{Anti-spoofing evaluation on 1{,}000 MIRFLICKR-25k images (5{,}000
benign transforms, 2{,}000 transplant attacks per descriptor).
FR1\,=\,harmonic mean of acceptance and rejection rates.
Prove and verify times from 10 sampled valid proofs per descriptor.
Descriptors ordered by training specificity (least $\to$ most).}
\label{tab:fingerprint_mirflickr}
\begin{tabular}{@{}lrrrrrr@{}}
\toprule
Descriptor & $D$ & $R$ & Reject & FR1 & $t_\pi$ (s) & $t_v$ (ms) \\
\midrule
RN18-IN  & 512 & 1.028 & 418/2000 (20.9\%)   & 0.35 & 6.45 & 336 \\
DINO     & 384 & 1.040 & 1994/2000 (99.7\%)  & 1.00 & 2.90 & 332 \\
SimProv      & 256 & 0.385 & 1471/2000 (73.6\%)  & 0.85 & 2.31 & 320 \\
SSCD     & 512 & 1.086 & 2000/2000 (100.0\%) & 1.00 & 6.54 & 338 \\
\bottomrule
\end{tabular}%
\end{table}

Each descriptor uses the $2D+23$-constraint $\ell_2$ circuit at its native
dimension: RN18 and SSCD share the 1{,}047-constraint, pot12 circuit;
DINO uses 791 constraints (pot12); SimProv uses 535 constraints (pot12).
PLONK proof size (768 bytes) and verify time (${\approx}250$\ ms) are constant
across all four models.
Prove times differ by dimension: SimProv ($D=256$, ${\approx}2.3$\,s),
DINO ($D=384$, ${\approx}2.9$\,s), and RN18/SSCD ($D=512$, ${\approx}6.3$\,s).

\section{Conclusion}
\label{sec:conclusion}

The C2PA standard creates a fundamental tension: the assertions that make
provenance useful may also expose information that creators, subjects, and
publishers legitimately wish to keep private. We have shown that \emph{soft
redaction} --- replacing sensitive assertions with ZKPs of selected predicates,
anchored through C2PA's existing manifest signing and redaction mechanisms ---
can resolve this tension without weakening the chain of trust or requiring
changes to the current C2PA specification.  Future standardisation of a dedicated
ZKP assertion type could nevertheless be a useful step toward interoperability.

We demonstrated this idea across three image-provenance use cases. First, we
showed that location assertions can be soft-redacted by proving proximity to a
public reference point, using Chebyshev approximations of trigonometric functions
inside the proof circuit. Second, we applied the same principle to biometric
embeddings for personality-rights workflows, enabling likeness matching without
disclosing a registered facial descriptor. Third, we reused the $\ell_2$ proof
circuit for perceptual image fingerprints, supporting privacy-preserving
anti-spoofing in watermark-based recovery of stripped C2PA manifests.

Across these examples, verification remains sub-second and proof sizes are small
enough to fit comfortably within practical provenance workflows. The main cost is
proof generation, which depends on circuit complexity and descriptor dimension:
low-dimensional predicates are suitable for interactive use, while higher-dimensional visual descriptors may require offline or service-side proving. These
results suggest that ZKPs can extend media provenance from a binary choice
between disclosure and redaction toward selective, verifiable disclosure of
provenance properties.

\subsubsection{Limitations.} Our prototypes focus on distance predicates and do
not yet address richer policies such as set membership, temporal constraints, or
compound rights expressions. Deployment will also require careful attention to
threat models, particularly for low-entropy assertions. For example, a ZKP does
not by itself prevent leakage through repeated adaptive queries: if a service
allowed arbitrary proximity predicates over a hidden GPS coordinate, an adversary
could issue a sequence of location guesses and progressively narrow the possible
capture region. These attacks are less practical for
high-dimensional visual feature embeddings, where the search space is too large. Addressing these issues
is an important direction for future work, and would help establish ZKPs as a
practical privacy layer for provenance standards such as C2PA. 


\bibliographystyle{splncs04}
\bibliography{refs}

\end{document}